\documentclass[intlimits,twoside,a4paper]{article}

\usepackage{amsmath,amssymb}
\usepackage{wrapfig}
\usepackage{graphicx}
\usepackage[T2A]{fontenc}
\usepackage[cp1251]{inputenc}
\usepackage{epstopdf}
\usepackage[eqsecnum]{cmpj2}

\issue{2015}{18}{2}{23801}
\doinumber{10.5488/CMP.18.23801}

\title
{Microphase transitions of block copolymer/homopolymer under shear flow}

\author[Y.~Guo \textsl{et al.}]{Y.~Guo, J.~Zhang, B.~Wang, H.~Wu, M.~Sun, J.~Pan}

\authorcopyright{Yu-qi Guo, Jin-jun Zhang, Bao-feng Wang, Hai-shun Wu, Min-na Sun, Jun-xing Pan, 2015}

\address{School of Chemistry and Materials Science, Shanxi Normal University, Linfen, 041004, China}

\date{Received July 29, 2014, in final form October 26, 2014}
\begin{document}

\maketitle

\begin{abstract}
Cell dynamics simulation is used to investigate the phase behavior of block copolymer/homopolymer mixture subjected to a steady shear flow. Phase transitions occur from transverse to parallel and then to perpendicular lamellar structure with an increase of shear rate and this is the result of interaction between the shear flow and the concentration fluctuation. Rheological properties, such as normal stress differences and shear viscosity, are all closely related with the direction of the lamellae. Furthermore, we specifically explore the phase behavior and the order parameter under weak and strong shear of two different initial states, and realize the importance of the thermal history. It is necessary to apply the shear field at the appropriate time if we want to get what we want. These results provide an easy method to create ordered, defect-free materials in experiment and engineering technology through imposing shear flow.


\pacs 83.50.Ax, 64.75.Yz, 83.80.Tc, 83.10.Tv
\end{abstract}

\section{Introduction}

One of the key targets in modern material science is preparation of highly ordered and controllable nano-structures. Spontaneously formed structures usually do not exhibit long-range order, and often contain a large number of defects, such as dislocations and disclinations. Thus, it is an important prerequisite that the emerging structures should be relatively defect-free. The long-range order can be induced by external guidance such as the shear field due to the inherent softness of polymers. The self-assembly phenomena of polymers under shear field analogous to those in a range of synthetic and biological systems can provide routes to the creation of novel materials. Furthermore, it is also easy, simple and operable to apply the shear field. Thus, this is a versatile means to obtain the long-range order and then create microstructures with potential applications in biomaterials, optics, and microelectronics. The use of shear has proven to be an excellent method for achieving long-range order during the past decades, and the phase behavior and rheological properties of polymeric materials subjected to shear flow also has made a remarkable progress in both experiments and simulations   \cite{key1,key2,key3,key4,key5,key6,key7, key8,key9,key10,key11}. Earlier, the study on block copolymer concentrated on experimental investigations, and most of them observed various alignments in a lamellar diblock copolymer by using TEM and SAXS, polystyrene-polyisoprene being the most representative. Parallel and perpendicular orientations were observed by Winey et~al. in 1993 \cite{key90} and Patel et al.  \cite{key100}  in 1995, respectively; transverse orientation was found by Gupta et~al. \cite{key110} and Zhang \cite{key12}, and it was defined as the one having a lamellae oriented perpendicular to both the shearing surfaces and the shearing direction. Then, Pinheiro and Winey observed mixed parallel-perpendicular morphologies in diblock copolymer systems at intermediate temperatures \cite{key13}. Afterwards, researchers theoretically predicted a phase transition of the lamellar \cite{key14,key15,key16,key17}, ring \cite{key18}, hexagonal cylinders \cite{key20}, diblock copolymer subjected to shear flow further. These predictions were proved by researchers using various numerical simulation methods, such as cell dynamics simulation (CDS) \cite{key22,key23,key24,key25,key26,key27,key28}, nonequilibrium molecular dynamics simulation (NEMD) \cite{key28,key29,key30,key31}, dissipative particle dynamics method (DPD) \cite{key32,key33,key34}, self-consistent field theory (SCFT) and lattice Boltzmann (LB) method \cite{key35}, Brownian dynamics (BD) \cite{key36}, mean-field approach (MFA) \cite{key37}, density functional theory (DFT) \cite{key308,key38}, Monte Carlo simulation (MC) \cite{key39}, molecular dynamics simulation (MD) \cite{key40} etc., Luo et~al. discovered that the parallel orientation is stable at low shear rate and the perpendicular orientation is stable at high shear rate in the hexagonal cylinder phase of asymmetric block copolymer, and explored the kinetic pathways based on the time-dependent Ginzburg-Landau approach \cite{key25}. Later on, the phase transition of gyroid phase \cite{key26}, and sphere phase \cite{key27,key308} of an asymmetric block copolymer under shear was also investigated. As for the symmetrical block copolymer, Lisal and  Brennan's simulation results  indicated that the perpendicular lamellar phase persists for all shear rates investigated, whereas the parallel lamellar phase is only stable at low shear rates, and it becomes unstable at high shear rates \cite{key34}. In addition, Guo studied the amphiphilic model system: its kinetics of the shear-induced isotropic-to-lamellar transition \cite{key31}, and the parallel-to-perpendicular orientation transition in the amphiphilic lamellar phase under shear flow \cite{key29,key30}. Fraser et~al.  investigated how shear flow affects the orientation of lamellar structures. They found out that for any finite shear rate, the entropy production and the average energy are the smallest if the lamellae's normal is perpendicular to the shear and the shear gradient direction, and this finding strongly suggests that the orientation corresponds to the global minimum for the bulk \cite{key40}. For all these cases, the shear flow plays an important role as a means for aligning the microscopic domains.

Even though the phase behavior of diblock copolymer systems under shear flow is relatively well understood, only a limited amount of information is available on the response of a complex polymer to shear. However, during the polymer processing, it is always doped with other substances in a block copolymer with the development of industrial synthesis technology, such as homopolymer, nanoparticle. The rheological behavior of binary mixtures of a polystyrene-polyisoprene-polystyrene (SIS) copolymers and homopolymer polystyrenes (PS) were studied by Baek and Han \cite{key41}. It is showed that the rheological behavior of mixtures is closely related to their morphological state, which in turn depends, among many other factors, on the ratio of the blends, molecular weight of the added homopolymer, and temperature. Further experiments also investigated the shear-induced phase behavior of block copolymer/homopolymer \cite{key42} and block copolymer/epoxy \cite{key403} blends. Later on, some researchers focused their attention on the phase behavior of block copolymer/nanoparticle \cite{key43,key44,key45,key46,key47,key48} composites under shear flow. Hence, He et~al. systematically investigated the shear-induced reorientations and phase transitions of symmetric diblock copolymer/nanorod nanocomposites subjected to a steady shear flow \cite{key47} and oscillatory shear flow \cite{key48} via DPD.

Consequently, it is really of great importance to explore the mysteries of block copolymer/homopolymer under external fields. Thus, the rest of this paper is devoted to the investigation of the phase behavior and rheological properties in a block copolymer/homopolymer mixture by using the cell dynamics simulation of time dependent Ginzbrug-Landau theory proposed by Oono and co-workers \cite{key49,key50,key51,key52,key53}. It is expected that this will  provide some guidelines for experimentalists. The other parts of this paper are organized as follows: section~2 is devoted to the description of the model and simulation method; section~3 is the numerical results and discussions; and finally, section~4 gives a brief conclusion of this work.

\section{Models and simulation methods}

We consider a block copolymer/homopolymer mixture rapidly quenched into the spinodal region, then subjected to a steady shear flow. Each copolymer chain is composed of monomers $A$ and $B$ with a short-range repulsive interaction between them. The interaction between $C$ monomer in a homopolymer and $B$ monomer in a copolymer is also assumed to be repulsive to each other. Hydrodynamic effects that prevail at the very late stage of phase separation in a polymer blend are neglected in the present model.

Several parameters are defined to describe the system. We take $\phi_{A0}$,  $\phi_{B0}$, and $\phi_{C0}$ as the average volume fractions of monomers $A$, $B$, and $C$. In the case of symmetric block copolymers which is considered here,  polymerization indices of the $A$ and $B$ blocks are the same; this is ensured by $\phi_{A0}=\phi_{B0}$. An important quantity, the composition ratio of homopolymers and copolymers which can be changed, is defined by
$f=f_{C0}\left/(f_{A0}+f_{B0})\right.$. In the process of phase separation, fluctuations are dominant, so we should investigate the local volume fractions of monomers $A$, $B$, and $C$. They are denoted, respectively, by
$\phi_A(x,y,z)$, $\phi_B(x,y,z)$, $\phi_C(x,y,z)$. Under the incompressibility condition, that is the total density $\phi_{A}(x,y,z)+\phi_{B}(x,y,z)+\phi_{C}(x,y,z)$ is constant, two of the local volume fractions are independent. Then, we take
$\phi(x,y,z)=\phi_{A}(x,y,z)-\phi_{B}(x,y,z)$, and $\psi(x,y,z)=\phi_{A}(x,y,z)+\phi_{B}(x,y,z)$ are independent variables used to characterize the structure ordering. The order parameter $\phi(x,y,z)$ gives the local concentration difference between the $A$ and $B$ monomers, the order parameter $\psi(x,y,z)$ describes the segregation of the homopolymer and the copolymer.

Here, we use a three-order-parameter model in \cite{key54}.  The free-energy functional of the system is given by
\begin{equation}
    F=F_{\mathrm{L}}+F_{\mathrm{S}}\,,\label{1}
\end{equation}
the long-range part $F_{\mathrm{L}}$ and the short-range part $F_{\mathrm{S}}$ are given by
\begin{equation}
    F_L=\frac{\alpha}{2} \iint {\rm d}{\textbf{\emph r}}{\rm d}{\textbf{\emph r}'}G({\textbf{\emph r}},{\textbf{\emph r}' })[\phi({\textbf{\emph r}})-\phi_0][\phi({\textbf{\emph r}'})-\phi_0],\label{2}
\end{equation}
and
\begin{equation}
    F_{\mathrm{S}}=\iint{\rm d} x {\rm d} y \left[\frac{c_1}{2}(\nabla\psi)^2+\frac{c_2}{2}(\nabla\phi)^2+\omega(\psi,\phi)\right],\label{3}
\end{equation}
respectively, where $\alpha$, $c_{1}$, $c_{2}$ are all positive constants. The long-range part is relatively simple, in which $G({\textbf{\emph r}},{\textbf{\emph r}'})$ is the Green's function defined by the equation
$-\nabla^{2}G({\textbf{\emph r}}, {\textbf{\emph r}'})=\delta({\textbf{\emph r}}-{\textbf{\emph r}'})$, while $\phi_0$ is the spatial averages of $\phi$. We should set $\phi_{0}=0$ in the case of symmetric copolymers. As for the short-range part, the $c_1$ and $c_2$ terms correspond to the surface tensions. The local interaction term $\omega(\psi,\phi)$ could be replaced by $\omega(\eta,\phi)$, where $\eta=\psi-\psi_{C}$ with $\psi_{C}$ being the volume fraction at the critical point of the macrophase separation and being a constant determined by the parameters of the system. According to Ito et~al., \cite{key54,key55}  we obtain
\begin{equation}
    \omega(\eta,\phi)=\nu_1(\eta)+\nu_2(\phi)+b_1\eta\phi-\frac{1}{2}b_2\eta\phi^2,\label{4}
\end{equation}
where the functions $\nu_1(\eta)$, $\nu_2(\phi)$ are assumed to be even with respect to the arguments, $b_1$,\,$b_2$ are all positive constants. Equation~\eqref{4} prescribes a minimal model of the short-range part of the free energy in our system. In the symmetric case, $b_1=(-\chi_{AC}+\chi_{BC})/2$, $b_2=1/(\psi_{C}^2N_A)$. It is now clear that the $b_1$ term mainly originates from the interaction between the $A$-$B$ copolymer and the $C$ monomer. If the repulsive interaction strength $\chi_{BC}$ between $B$ and $C$ is large enough, $b_1$ is positive. $b_2$ term represents the fact that the microphase separation should occur only in the copolymer-rich phase.

In terms of the free energy functional in equations~\eqref{1}--\eqref{4}, we obtain a set of two coupled equations
\begin{equation}
      \frac{\partial\,\eta}{\partial\,t}=M_\eta\,\nabla^2\frac{\delta\,F(\eta,\phi)}{\delta\eta}-{\textbf{\emph v}}\cdot\nabla\,\eta({\textbf{\emph r}},t), \label{5}
\end{equation}
\begin{equation}
    \frac{\partial\,\phi}{\partial\,t}=M_\phi\,\nabla^2\frac{\delta\,F(\eta,\phi)}{\delta\phi}-{\textbf{\emph v}}\cdot\nabla\,\phi({\textbf{\emph r}},t), \label{6}
\end{equation}
where $M_\eta$ and $M_\phi$ are transport coefficients, $\textbf{\emph v}$ is an external velocity field describing the shear flow profile
\begin{equation}
  \textbf{\emph v}=\dot{\gamma}y{\textbf{\emph e}}_x\,, \label{7}
\end{equation}
where $\dot{\gamma}$ is the shear rate and ${\textbf{\emph e}}_x$ is the unit vector in the $x$ direction.
Numerical solutions of the above model system can be carried out in an $L_x \times L_y \times L_z$ three-dimensional cubic lattice, by using the cell dynamics simulation (CDS) approach, mainly because it is a fast method to simulate kinetic processes in phase separating systems of large sizes. The order parameters for each cell are $\eta({\textbf{\emph n} },t)$, $\psi({\textbf{\emph n}},t)$, where ${\textbf{\emph n}}=(n_x,n_y,n_z)$ is the lattice position and $n_x$, $n_y$, and $n_z$ are integers between 1 and $L$. The CDS equations corresponding to equations~\eqref{5}--\eqref{6}, in their space-time discretized form, are written as follows:
\begin{eqnarray}
  \eta(x,y,z,t+\Delta{t})&=&\eta(x,y,z,t)+M_\eta\left(\langle\langle I_\eta\rangle\rangle-I_\eta\right)-\frac{1}{2}\dot{\gamma}y \left[\eta(x+1,y,z,t)-\eta(x-1,y,z,t)\right],    \label{8}
\\
  \phi(x,y,z,t+\Delta{t})&=&\phi(x,y,z,t)+M_\phi\left(\langle\langle I_\phi\rangle\rangle-I_\phi\right)-\frac{1}{2}\dot{\gamma}y \left[\phi(x+1,y,z,t)-\phi(x-1,y,z,t)\right],    \label{9}
\end{eqnarray}
where
\begin{eqnarray}
    I_\eta&=&-D_1\left(\langle\langle{\eta}\rangle\rangle-\eta\right)-A_\eta\tanh\eta+\eta+b_1\phi-\frac{1}{2}b_2\eta\phi,\label{10}
\\    I_\phi&=&-D_2\left(\langle\langle{\phi}\rangle\rangle-\phi\right)-A_\phi\tanh\phi+\phi+b_1\eta-b_2\eta\phi,\label{11}
\end{eqnarray}
and
\begin{equation}
    \langle\langle{x}\rangle\rangle=\frac{6}{80}\sum_N{x}+\frac{3}{80}\sum_{NN}{x}+\frac{1}{80}\sum_{NNN}{x},\label{12}
\end{equation}
the subscripts $N$, $NN$, and $NNN$ stand for the nearest-neighbor, the next-nearest neighbor, and the next-next-nearest neighbor cells, respectively \cite{key52,key53}.  To perform the numerical operations in a cell dynamics system, in general, the lattice size ($\Delta x$,$\Delta y$ or $\Delta z$) and the time step ($\Delta t$) are all set to unity.

We choose $x$-axis as the flow direction, $y$-axis as the velocity gradient direction and $z$-axis as the vorticity axis. A shear periodic boundary condition proposed by Ohta et~al. \cite{key56,key57,key58,key59} should be applied to $x$ direction, and the periodic boundary conditions are applied in the $y$ and $z$ direction. With the shear strain $\gamma$, this boundary condition is written as follows:
\begin{equation}
   \eta(n_x,n_y,n_z,t)=\eta\left[ n_x+N_{x}L+\gamma(t)N_{y}L,n_y+N_{y}L,n_z+N_{z}L \right],  \label{13}
\end{equation}
\begin{equation}
    \phi(n_x,n_y,n_z,t)=\phi \left[ n_x+N_{x}L+\gamma(t)N_{y}L,n_y+N_{y}L,n_z+N_{z}L\right],\label{14}
\end{equation}
where $N_x$, $N_y$, and $N_z$ are arbitrary integers.

For the three-Dimensional space, the first and the second normal stress differences $N_1$ and $N_2$, are defined as \cite{key3,key60}
\begin{equation}
    N_1=-\int\frac{{\rm d}{\textbf{\emph k}}}{(2\pi)^3}\left[k_x^2-k_y^2\right]S({\textbf{\emph k}},t),\label{15}
\end{equation}
\begin{equation}
    N_2=-\int\frac{{\rm d} {{\textbf{\emph k}}}}{(2\pi)^3}\left[k_y^2-k_z^2\right]S({\textbf{\emph k}},t),\label{16}
\end{equation}
and the excess viscosity defined as \cite{key61,key62,key63}
\begin{equation}
    \Delta\eta=-\frac{1}{\gamma}\int\frac{{\rm d}{\textbf{\emph k}}}{(2\pi)^3}k_xk_yS({\textbf{\emph k}},t).\label{17}
\end{equation}

Our simulations are carried out on $L_x \times L_y \times L_z=64\times64\times64$ cubic lattice, and the parameters are chosen to be $A_\eta=1.3$, $A_\phi=1.1$, $D_1=1.0$, $D_2=0.5$, $b_1=0.10$, $b_2=0.02$, and $M_\eta=M_\phi=1$, the initial distributions of $\phi$ and $\eta$ are specified by random uniform distributions in the range $[-0.01,0.01]$, in accordance with the previous work  \cite{key64,key65,key66}.  In this paper, all parameters are scaled, so all of them are dimensionless \cite{key49}.  In the present simulation work, $\alpha=0.02$ and the formula is $\alpha={12}/{[N^2 f_{\mathrm{b}}(1- f_{\mathrm{b}})]}$ ($f_{\mathrm{b}}$ denotes a block ratio) \cite{key67}.

\section{Numerical results and discussion}

\begin{figure}[htb]
\vspace{10mm}
\centerline{
\includegraphics[width=0.55\textwidth]{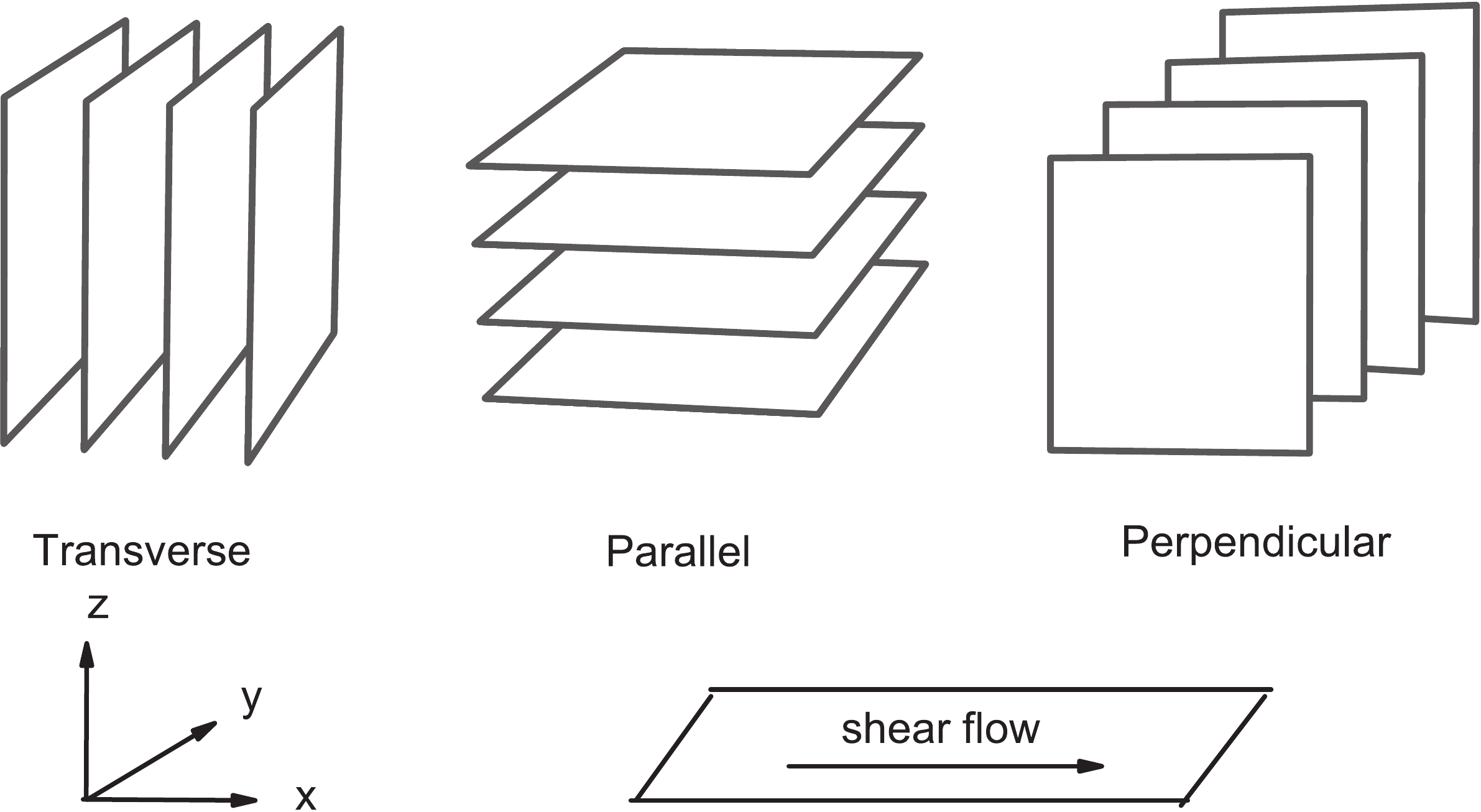}
}
\vspace{5mm}
\caption{Schematic representation of the geometry considered including the shear direction and three different lamellar orientations discussed in the text.} \label{fig1}
\end{figure}

\begin{figure}[!h]
\vspace{5mm}
\centerline{
\includegraphics[width=0.65\textwidth]{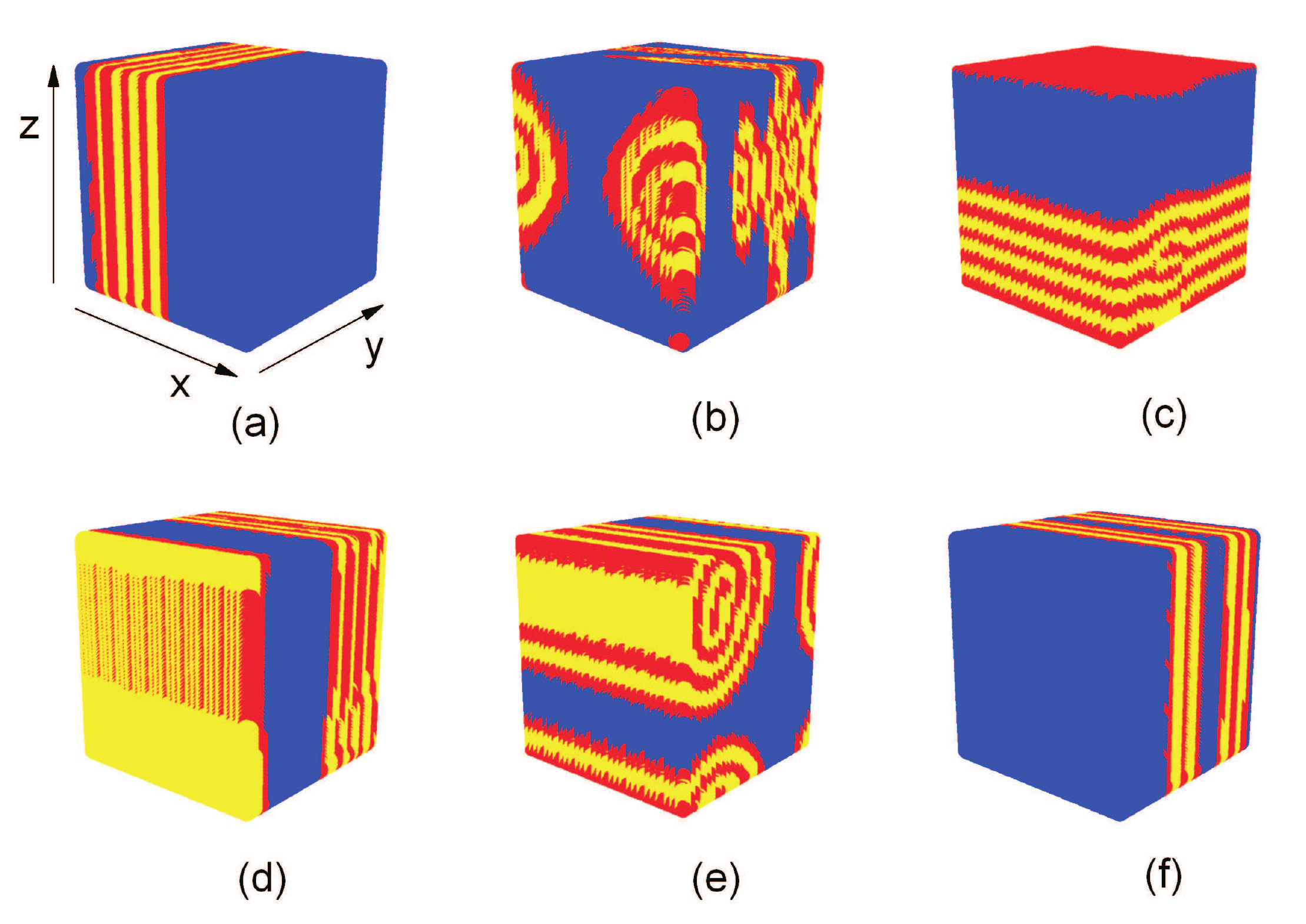}
}
\caption{(Color online) Block copolymer/homopolymer under different shear rates, $\dot{\gamma}=0.0,0.000001$, 0.000005,
0.000006, 0.0006, and 0.001 in (a)--(f). Phase $A$ is represented by the red regions, phase $B$ by the yellow regions, phase $C$ by the blue regions. The arrows in (a) show the direction of the $x$-axis, $y$-axis, $z$-axis, respectively.} \label{fig2}
\end{figure}

In order to investigate the phase behavior and rheological properties of the block copolymer and homopolymer mixture under steady shear, CDS simulation is performed on the systems. The shear direction and the schematic of three different lamellar orientations are shown in figure~\ref{fig1}, and they will be mentioned in the following discussion. Under zero-shear, we obtain abundant morphologies with different composition ratios of copolymer and homopolymer that are not displayed in this section. They conclude transverse, parallel, perpendicular lamellar structure and the concentric cylindrical structure that the block copolymer is surrounded by homopolymer when the homopolymer is in majority. Next, we want to know the structures corresponding to different shear rates relative to the transverse lamellae under zero-shear. Here, $f=55/45$, all other conditions are the same except the shear rate. It means that the initial state at all shear rates includes a zero-shear and is disordered. The imposed shear rates $\dot{\gamma}$ range from 0 to 0.001 and we show a part of them in figure~\ref{fig2}.


The ordered transverse lamellar structure mentioned above under zero-shear is shown in figure~\ref{fig2}~(a). With an increase of the shear rate, the chain stretches along the $x$-axis direction until it stretches completely. Owing to a different intensity of the shear flows, we obtain  different domain morphologies. The structure shows that the width of the domain is larger than the others corresponding to the shear rate $\dot{\gamma}=0.000001$ in figure~\ref{fig2}~(b), because the shear rate is too small to stretch entirely.
We need to consider the orientations of the model after the chain stretches completely along $x$-axis at an appropriate shear rate. A relatively weak shear more strongly suppresses the concentration fluctuations along the shear gradient direction. Thus, a perfect parallel phase is formed at the shear rate of  $\dot{\gamma}=0.000005$ in figure~\ref{fig2}~(c). It is interesting that when we slightly increase the shear rate, the model transforms into a perpendicular lamellae with some defect as shown in figure~\ref{fig2}~(d).
Then, continuously increasing the shear rate to an appropriate value, we get a concentric cylindrical structure where the block copolymer is surrounded by homopolymer in figure~\ref{fig2}~(e). Finally, at a higher shear rate, since the shear flow more strongly suppresses  the concentration fluctuation along the vorticity direction, the model turns to the ordered perpendicular phase, as can be seen in figure~\ref{fig2}~(f). Besides this, the phenomenon also demonstrates that a low shear rate is beneficial for the formation of a parallel lamellar structure; and the lamellar phase preferentially adopts a perpendicular orientation at a high shear rate.

As concerns the phenomenon under a simple steady shear flow, we can explain that it is a result of a competition between the shear flow and the concentration fluctuation. The imposition of a shear flow suppresses the concentration fluctuation inordinately. Under a relatively weaker shear, the fluctuation effect is larger than the field effect, hence the chain cannot completely stretch along the shear direction. However, we have to admit that a weak shear also suppresses the concentration fluctuation to some extent, and it suppresses the fluctuation more strongly along the velocity gradient axis, i.e., $y$ axis. This is the main cause of preferential adoption of the parallel orientation under a weak shear flow. The higher the shear rate is, the greater the field effect is. Under a stronger shear flow, the field effect is predominant, and it intensively suppresses the concentration fluctuation. Moreover, a stronger shear more strongly suppresses the fluctuation along the vorticity direction, i.e., $z$ axis, thus, it transforms into the ordered perpendicular lamellar structure when the shear rate increases to a certain degree. This is in agreement with the previous study of the block copolymer under shear flow by K. A. Koppi et al. in experiments, \cite{key68} and by Igor Rychkov  in theoretical studies,  \cite{key28}, respectively.

\begin{wrapfigure}{o}{0.5\textwidth}
\centerline{\includegraphics[width=0.37\textwidth]{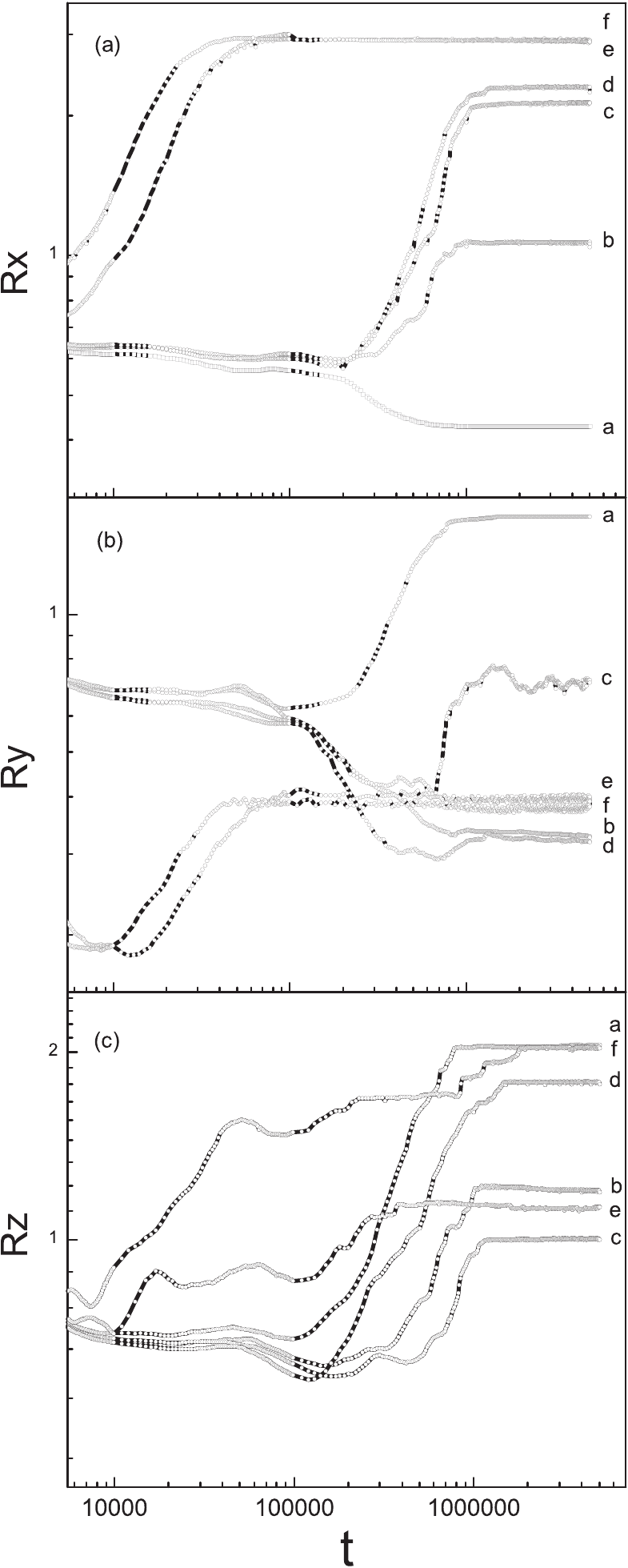}}
\caption{Time evolution of the characteristic sizes of microdomains $R_x(t)$, $R_y(t)$ and $R_z(t)$ for different shear rates, $\dot{\gamma}=0.0$, 0.000001, 0.000005, 0.000006, 0.0006, and 0.001 in curves a--f.} \label{fig3}
\end{wrapfigure}

To consider this case in greater depth, we would numerically calculate the domain sizes $R_i(t)$ $(i=x,\,y, \text{or}\,z)$ in the $x$, $y$, or $z$ direction as a function of time. The domain sizes $R_i(t)$ can be derived from the inverse of the first moment of the structure factor $S(\textbf{k},t)$ as follows:
\begin{equation}
    R_i(t)=2\pi/\langle k_i(t)\rangle,\label{18}
\end{equation}
where
\begin{equation}
    \langle|k_i(t)|\rangle=\int{\rm d} {\textbf{k}}k_iS(\textbf{k},t)\left/\int {\rm d}{\textbf{k}}S({\textbf{k}},t)\right..\label{19}
\end{equation}
In fact, the structure factor $S(\textbf{k},t)$ is determined by the Fourier component of the spatial concentration
distribution \cite{key49,key50,key51,key52}. It is defined by $S(\textbf{k},t)=\langle|\phi(\textbf{k},t)|\rangle^2 $ \cite{key53}. Figure~\ref{fig3} shows the time evolution of the microdomain sizes $R_i(t)$ in the $x$, $y$, and $z$ directions as a function of time in the double-logarithmic plots. The results are averaged over ten independent runs.

From figure~\ref{fig3}~(a), we can see that the microdomain size $R_x$ in equilibrium gradually increases (from curve $a$ to curve $e$) as the shear rate increases, and reaches a maximum value in curve $e$ which corresponds to the shear rate $\dot{\gamma}=
0.0006$ in figure~\ref{fig2}~(e). Continuously increasing the shear rate, $R_x$ in equilibrium remains constant (from curve $e$ to $f$). This fact shows that the growth of the microdomain size takes place along the $x$-axis (i.e., the shear flow) with an increase of the shear rate. We also see that the difference between the curves $e$ and $f$ is that a relatively short time is needed to reach the equilibrium at a higher shear rate. Although the change of the microdomain size $R_y$ in figure~\ref{fig3}~(b) and $R_z$ in figure~\ref{fig3}~(c) from the curve $a$ to curve $f$ is more complex, it is consistent with the figure~\ref{fig2} completely. We clearly see a steep increase from curves $b$ to $c$ and a sharp decrease from curves $c$ to $d$ in figure~\ref{fig3}~(b); on the contrary, we see a sudden decrease from curves $b$ to $c$ and a rapid increase from curves $c$ to $d$ in figure~\ref{fig3}~(c). It means that we obtain the order parallel phase under the shear rate of curve $c$, and the defective perpendicular phase under the shear rate of curve $d$. Moreover, $R_y(c)\gg R_y(f)$ and $R_z(c)\ll R_z(f)$ confirm that the lamellae adopt a parallel orientation corresponding to figure~\ref{fig2}~(c) and perpendicular orientation corresponding to figure~\ref{fig2}~(f), respectively. From figure~\ref{fig3} we also see that the microdomain size grows slower at an early stage while it changes apparently in the middle stage, then it is stable at the later stage. With respect to the lower shear rate, the higher shear makes the microdomain grow faster, until achieving a constant value.

\begin{figure}[htb]
\centerline{
\includegraphics[width=0.75\textwidth]{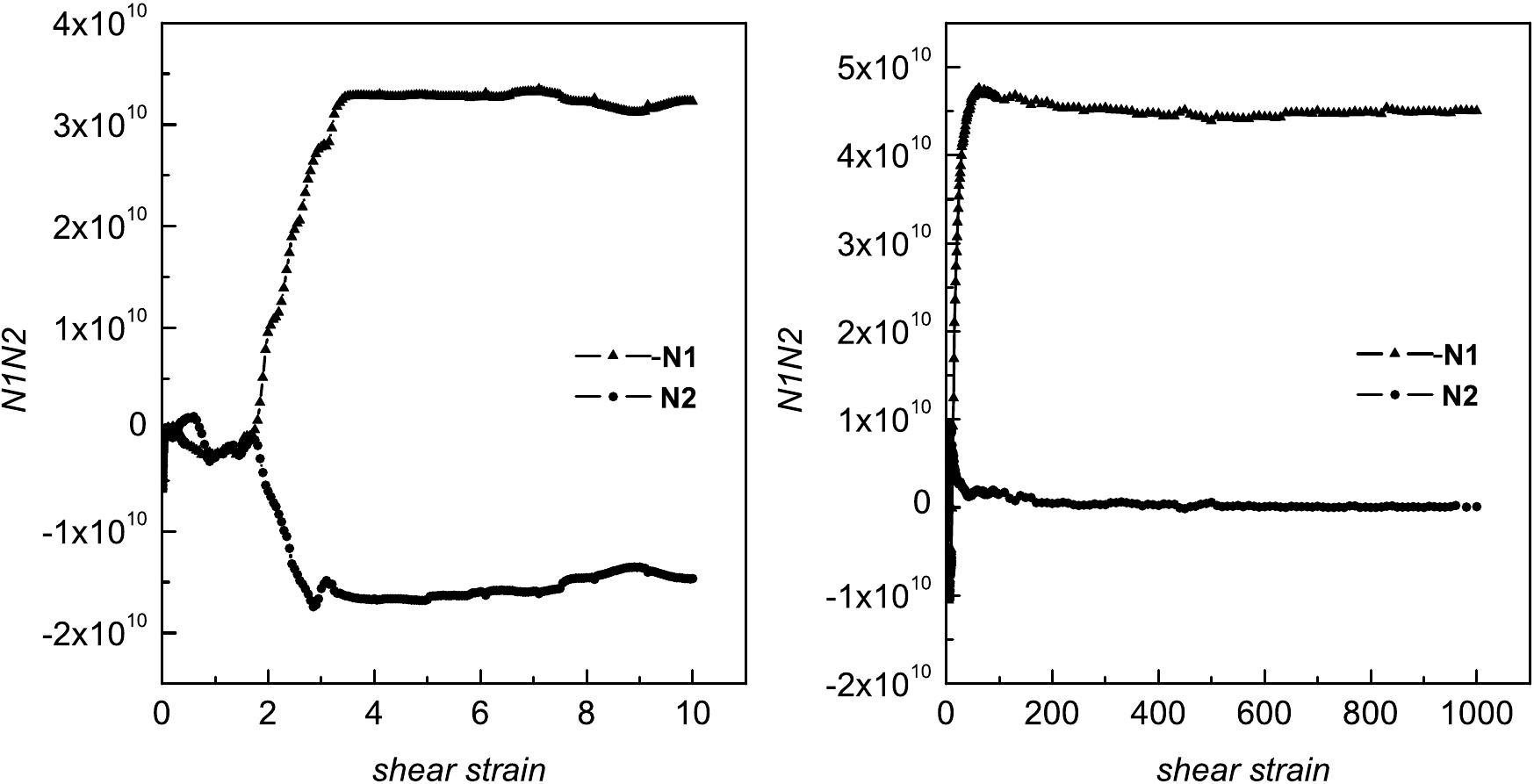}
}
\caption{The first and the second normal stress differences $N_1$ and $N_2$ as a function of the shear strain for (a) $\dot{\gamma}=0.000005$; (b) $\dot{\gamma}=0.001$.} \label{fig4}
\end{figure}

\begin{figure}[htb]
\centerline{
\includegraphics[width=0.55\textwidth]{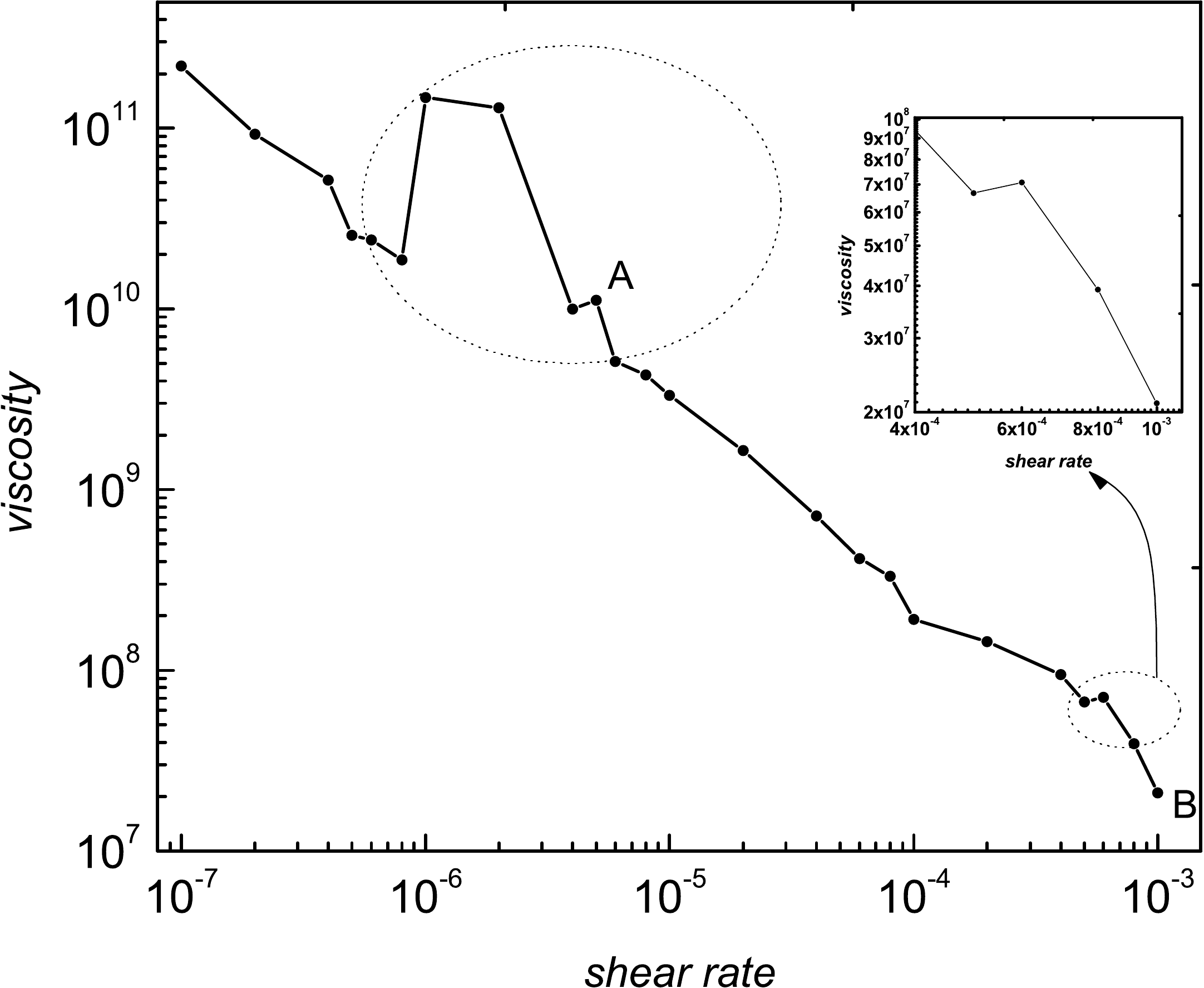}
}
\caption{The shear viscosity $\eta$ as a function of the shear rate $\dot{\gamma}$ for the block copolymer and homopolymer.} \label{fig5}
\end{figure}

Owing to the dual properties including elastic deformation and viscous flow, the polymer liquid often shows more complex properties under the external force. We know that it is meaningless to study the absolute value of the normal stress component
individually, and the difference value of the normal component always stays the same, so we can use the first and the second normal stress differences value to describe the stress state. For a simple shear flow, we use the shear viscosity to describe the viscosity of a system.

Figure~\ref{fig4} shows the first and the second normal stress differences $N_1$ and $N_2$ as a function of shear strain under weak and strong shear, respectively. We can clearly see that $N_1$ and $N_2$ are sensitively dependent on the direction of the lamellae. Figure~\ref{fig4}~(a) corresponds to the case of forming a parallel lamellar structure. It indicates that $N_1>0$, and it increases apparently along with an increase of the shear strain until achieving the constant value. On the contrary, $N_2<0$, and it decreases apparently along with an increase of the shear strain until reaching a constant value. Figure~\ref{fig4}~(b) corresponds to the case of forming perpendicular lamellar structure. Compared with figure~\ref{fig4}~(a), it indicates that $N_1>N_2>0$, and they increase along with an increase of the shear strain until reaching  a constant value.

As we know, the original formulas of the first and the second normal stress differences are $N_1=\sigma_{xx}-\sigma_{yy}$, and $N_2=\sigma_{yy}-\sigma_{zz}$. $\sigma_{xx}$,\, $\sigma_{yy}$,\, $\sigma_{zz}$ are the normal principal stresses, and express the tension in $x$, \,$y$, and $z$ directions, respectively. Here, we should consider the pressure that is imposed on the system in $x$, \,$y$, and $z$ directions in order to correspond to the formulas~\eqref{15} and \eqref{16}. Therefore, we have a good explanation that $N_1$ and $N_2$ are sensitively dependent on the direction of the lamellae. We can take no account of the stress of the $x$-axis direction since $x$-axis is the direction of the shear flow, and it is mainly affected by the tangential stress. For the case of forming a parallel alignment, $N_2<0$, it means that the pressure in $y$-axis direction is smaller than that in $z$-axis direction; for the case of forming a perpendicular alignment, $N_2>0$, it means that the pressure in $y$-axis direction is larger than that in $z$-axis direction. This verifies that the low shear rate is beneficial to the formation of a  parallel lamellar structure; and the lamellar phase preferentially adopts the perpendicular orientation at a high shear rate.

Figure~\ref{fig5} shows the calculated shear viscosity of the block copolymer/homopolymer as a function of the shear rate. Generally speaking, the shear viscosity gradually decreases with an increase of the shear rate, corresponding to shear thinning, in which the layer width is slightly decreased so that the transverse lamellae can adopt parallel or perpendicular orientation \cite{key34}.
However, in addition to the shear thinning, there also exists a shear thickening in figure~\ref{fig5}, corresponding to the local convexes. The first larger local convex corresponds to the figure~\ref{fig2}~(b), the width of the domain is larger than the others; and the second local convex corresponds to the concentric cylindrical structure in figure~\ref{fig2}~(e) for the same reason as the first one. Points A and B in figure~\ref{fig5} are the viscosity under the shear rates 0.000005 and 0.001, respectively. This indicates that the viscosity of the perpendicular phase is by far smaller than the parallel phase.

\begin{figure}[htb]
\centerline{
\includegraphics[width=0.65\textwidth]{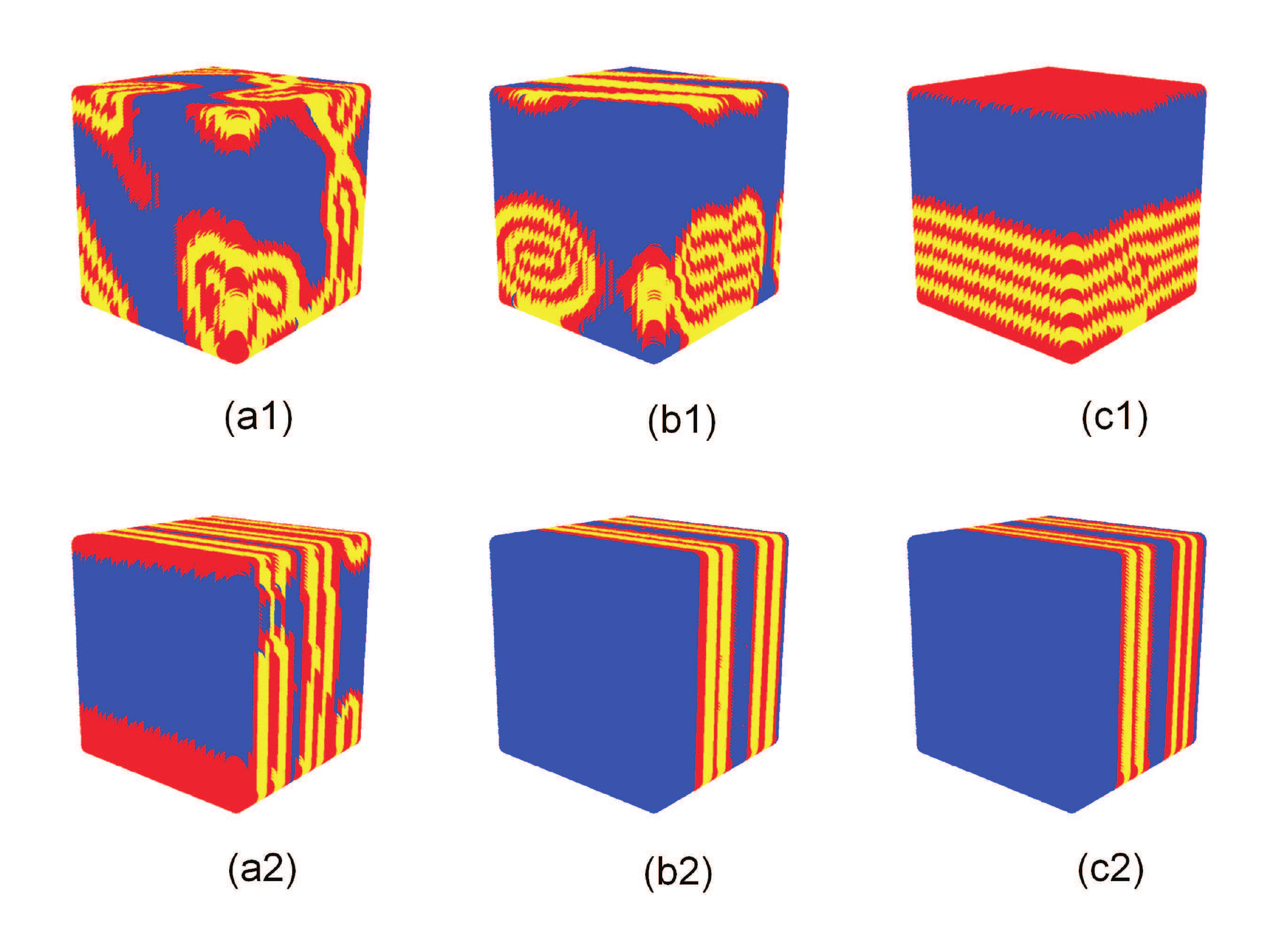}
}
\caption{(Color online) Simulation for block copolymer/homopolymer at different times corresponding to the weak and strong shear of $\dot{\gamma}=0.000005$ and $\dot{\gamma}=0.001$. The initial state is disordered. (ai): $t=50000$, (bi): $t=500000$, (ci): $t=5000000$. Other details are the same as in figure~2.} \label{fig6}
\end{figure}
\begin{figure}[htb]
\centerline{
\includegraphics[width=0.63\textwidth]{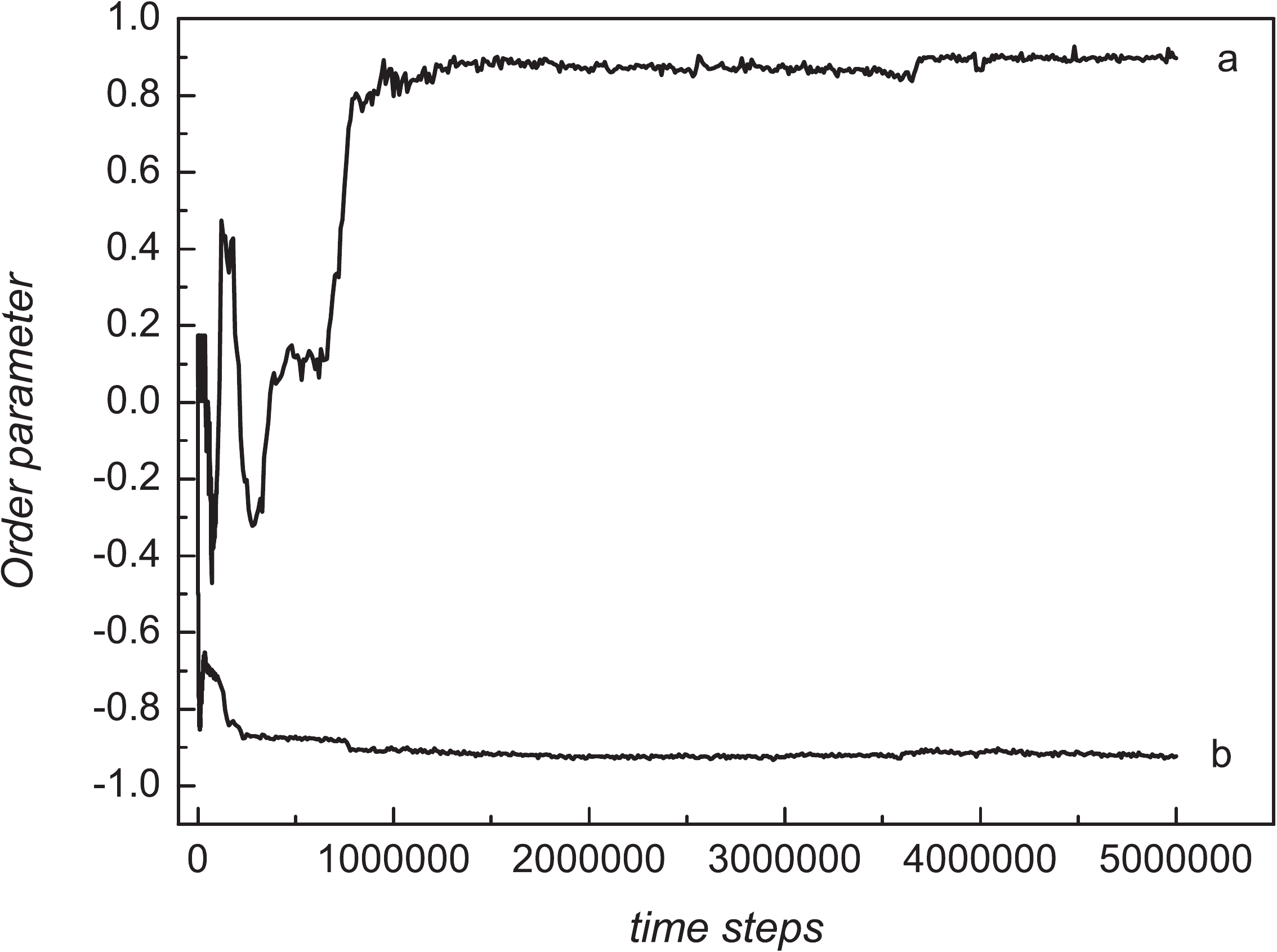}
}
\caption{The order parameter S of block copolymer/hopolymer as a function of time for weak and strong shear, corresponds to figure~6. a: $\dot{\gamma}=0.000005$ and b: $\dot{\gamma}=0.001$.} \label{fig7}
\end{figure}

Moreover, in order to monitor the processes of forming parallel and perpendicular lamellar structure in detail, we show the simulation snapshots at different times for $\dot{\gamma}=0.000005$ and $\dot{\gamma}=0.001$ in figure~\ref{fig6}. In other words, the two different shear rates represent the weak and the strong shear mentioned before.

We simulate the occurrence of the parallel lamellar structure under weak shear flow in figures~\ref{fig6}~(a1)--(c1). As could be expected, it is an irregular structure in the initial period of time, such as the structure in figure~\ref{fig6}~(a1). With the time evolution, the domain morphology begins to tend  to the parallel lamellae in figure~\ref{fig6}~(b1). Then, the domain morphology gradually evolves into the parallel phase with some defects. Due to the shear flow, the defects become fewer and an ordered parallel lamellar structure appears in figure~\ref{fig6}~(c1). For comparison, we also consider the occurrence of a perpendicular lamellar structure under a strong shear flow. As shown in figures~\ref{fig6}~(a2)--(c2), the formation of lamellae phase only needs a very short time. At $t=50000$, the system has formed a perpendicular lamellar structure with more defects. Moreover, compared with the disordered structure in figure~\ref{fig6}~(b1), it has formed a perfect perpendicular phase in figure~\ref{fig6}~(b2) because the shear field is so strong that it has an overwhelming superiority. Comparatively speaking, the process  of forming an ordered parallel phase is very slow because the  external field is weaker and  it has no advantage in competition with the concentration fluctuation.

Meanwhile, to verify the result further, we calculate the orientational order parameter $S$ of the block copolymer/homopolymer under a weak and strong shear flow,
\begin{eqnarray}
  S=\langle2\cos^{2}\theta-1\rangle,\label{20}
\end{eqnarray}
here, we choose to calculate the order parameter of $yz$ plane. Thus, $\theta$ is the angle between the unit normal vector of the lamellae in $yz$ plane and the unit normal vector in the velocity gradient axis \cite{key69,key70}. $S$ is the order parameter, $S=0$ represents the completely disordered state; $S=1$ means a completely orientational phase parallel to the velocity gradient direction; and $S=-1$ means another completely oriented phase perpendicular to the velocity gradient direction. Figure~\ref{fig7} displays the order parameter at two typical shear rates, $\dot{\gamma}=0.000005$ and $\dot{\gamma}=0.001$.

As depicted in figure~\ref{fig7}, the initial value of the order parameter $S$ in two cases is around zero. This means that the original state of the system is completely irregular. For the case of $\dot{\gamma}=0.000005$ in curve $a$, the order parameter is in a mess before the time steps $t=750000$, so the microphase structures are disordered in figures~\ref{fig6}~(a1)--(b1). With time evolution, the order parameter $S$ increases slowly because the shear flow gradually plays a certain role. Finally it reaches a certain value which suggests that the system forms an ordered structure, i.e., a parallel lamellar structure, which corresponds to figure~\ref{fig6}~(c1). In comparison with the above result, the order parameter $S$ in curve $b$ decreases quickly with an increase of time under strong shear, i.e., \,$\dot{\gamma}=0.001$, and it drops the lowest in a remarkably short time. It corresponds to figures~\ref{fig6}~(a2)--(c2), where the morphology quickly reaches the final equilibrium structure. In addition, the order parameter $S$ has almost no change at the late stage, so the microphase structures that we finally obtain  are stable.
\begin{figure}[htb]
\centerline{
\includegraphics[width=0.65\textwidth]{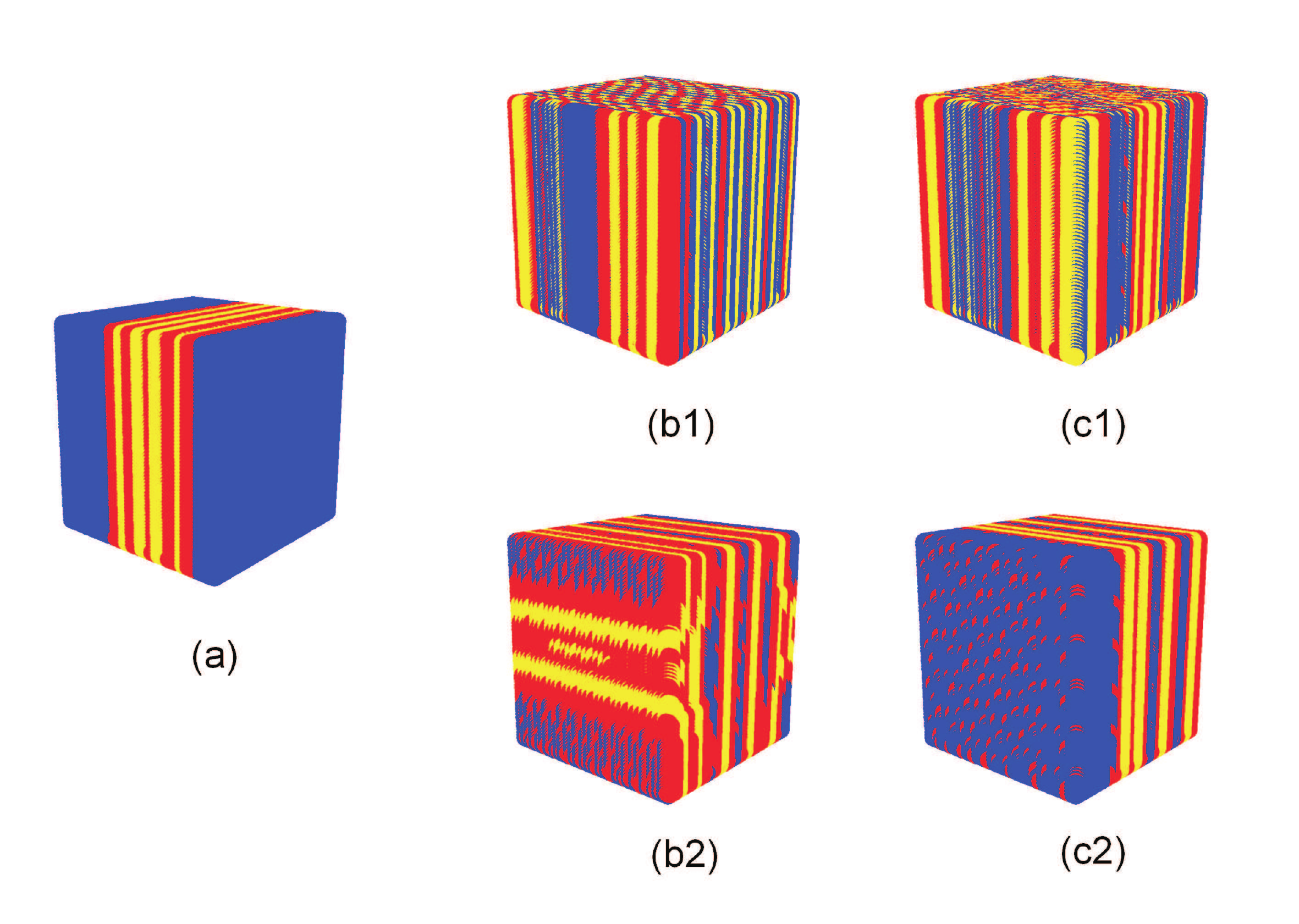}
}
\caption{(Color online) Simulation for block copolymer/homopolymer at different times under a weak and strong shear of $\dot{\gamma}=0.000005$ and  $\dot{\gamma}=0.001$. The initial state is ordered. (a): $t=0$, (b1): $t=1500000$, (c1): $t=3500000$, (b2): $t=50000$, (c2): $t=3500000$. Other details are the same as in figure~2.} \label{fig8}
\end{figure}

Next, we impose these two typical shear rates on the equilibrium system which arise from the self-assembly of block copolymer/homopolymer under zero-shear and explore the simulation snapshots at different times in figure~\ref{fig8}. The equilibrium phase under zero-shear indicates that the initial state is ordered as shown in figure~\ref{fig8}~(a). We see that the domain structure merely tilts along with the shear flow under a weak shear flow as shown in figures~\ref{fig8}~(b1)--(c1). At the same time, a strong shear flow makes the ordered lamellar structure break up quickly and reform the perpendicular lamellar structure with some defects in figure~\ref{fig8}~(b2). The defects are eliminated with the time evolution, and it transformed to the perpendicular phase with some perforated lamellae in figure~\ref{fig8}~(c2).
\begin{figure}[htb]
\centerline{
\includegraphics[width=0.63\textwidth]{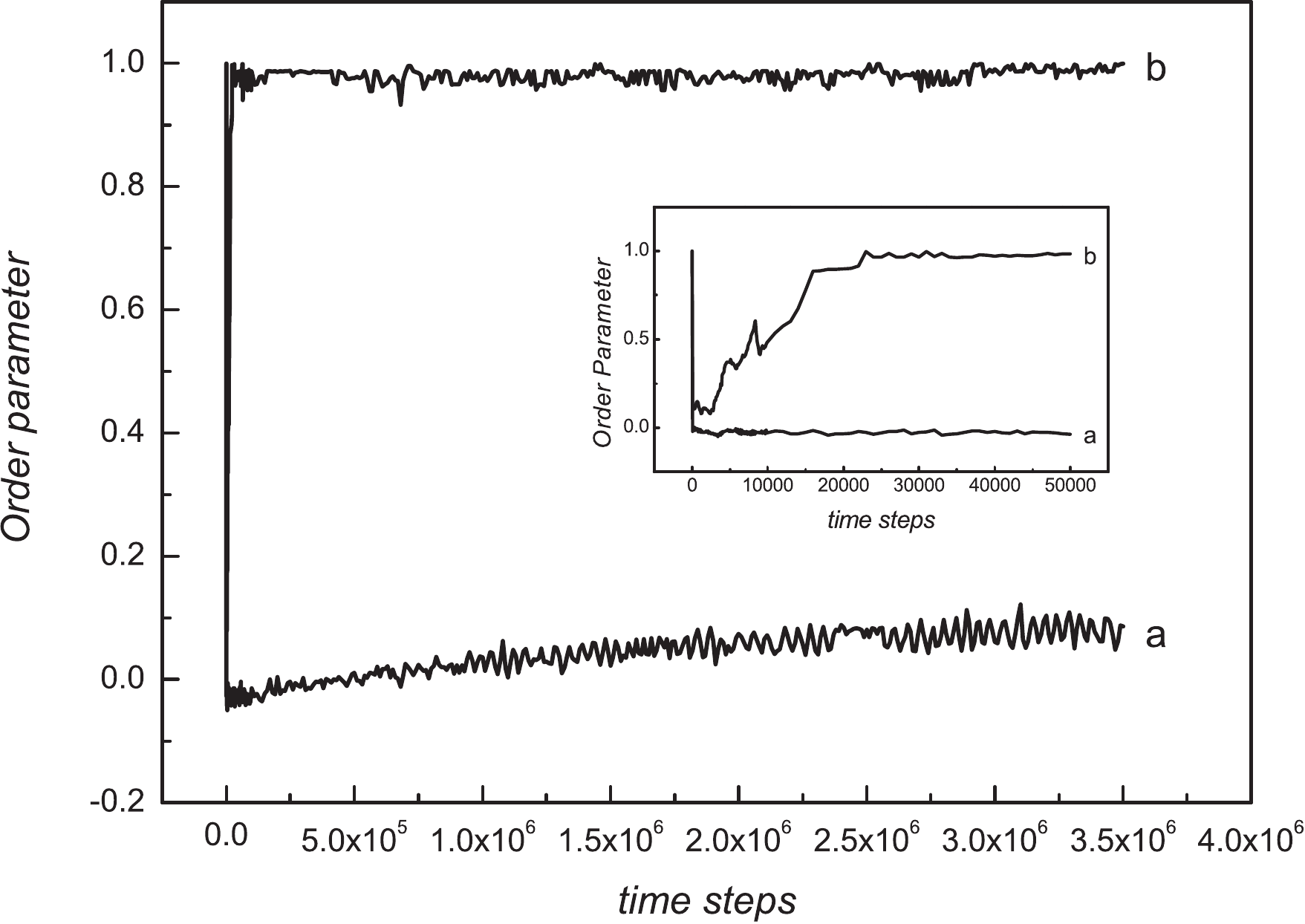}
}
\caption{The order parameter S of block copolymer/hopolymer as a function of time for weak and strong shear, corresponds to figure~6. a: $\dot{\gamma}=0.000005$ and b: $\dot{\gamma}=0.001$.} \label{fig9}
\end{figure}

Figure~\ref{fig9} shows the order parameter corresponding to figure~\ref{fig8}. We choose to calculate the order parameter of $xy$ plane. Thus, $\theta$ is the angle between the unit normal vector of the lamellae in $xy$ plane and the unit normal vector in the shear flow axis. $S=0$ represents a completely disordered state, $S=1$ means a completely orientational phase parallel to the shear flow axis. In the case of $\dot{\gamma}=0.000005$, $S$ quickly decreases to reach its minimum, and then slightly increases around $S=0$. This indicates that although the shear flow is too weak to break up the ordered transverse, the weak shear could make the ordered phase turn to a disordered phase tilting to the x-axis. In the case of $\dot{\gamma}=0.001$, $S$ also quickly decreases to reach its minimum around 0, but then it increases to a maximum around 1 in a short time and keeps a stable state during the next time. This result is consistent with the structural evolution shown in figures~\ref{fig8}~(a)--(c2).

From the above result, we can give the following explanations. When the initial state is ordered, the weak shear that we imposed is too weak to break up the ordered transverse phase. Thus, it cannot transform to another ordered structure. However, as figures~\ref{fig6}~(a1)--(c1) mentioned, although the shear flow is also weaker, the initial structure is disordered, it is easy to form another ordered structure with time evolution. In addition, compared with the transformation from a transverse to a parallel phase of the block copolymer/nanorod composites \cite{key47}, the whole interfacial tension increases when doped with the homopolymer, so the interfacial tension of the block copolymer/homopolymer is larger than the block copolymer/nanorod, and then the former's order-disorder transition temperature ($T_{\mathrm{ODT}}$) is higher than the latter's. Thereby, it makes the order-order transition more difficult under a weak shear. In other words, this is the effect of the homopolymer. However, under a strong shear, the domain morphologies all finally turn to the perpendicular phase in the above mentioned cases including block copolymer/nanorod composites. This is because  the shear flow is strong enough and absolutely superior, notwithstanding whether the initial state is ordered or disordered, and whether the system is block copolymer/nanorod or block copolymer/homopolymer.

From another perspective, the thermal history cannot be ignored owing to the initial structure is the equilibrium phase under zero-shear. In fact, this is equivalent to the fact that the imposed shear flow after the phase is separated completely. So, we consider that thermal history also has a certain effect on the difference between figure~\ref{fig6} and figure~\ref{fig8}. Seung Su Kim \cite{key71} and M. L. Cerrada \cite{key72} have studied the effect of thermal history on the phase behavior of  polymeric materials, and it showed that some contradictory results in the references are likely to be caused by the different thermal history of a sample.

\section{Conclusions}

In this paper, the cell dynamics simulation is used to investigate the phase behavior and rheological properties of a block copolymer and homopolymer mixture subjected to a steady shear flow. Different morphologies corresponding to different shear rates, and the phase transitions occur from the transverse to parallel and then to perpendicular lamellar structure with an increase of the shear rate. It is a result of the competition the between shear flow and the concentration fluctuation. The weak shear more strongly suppresses the concentration fluctuation along the velocity gradient axis, while a strong shear strongly suppresses it along the vorticity axis, which  is caused by the formation of a parallel and perpendicular phase under a weak and strong shear, respectively.  This is in agreement with the previous work on block copolymer which was studied by K. A. Koppi et al. in experiments \cite{key68}, and by Rychkov Igor in theory  \cite{key28}. Rheological properties being considered show that the first and second normal stress differences are sensitively dependent on the direction of the lamellae. The shear viscosity gradually decreases with an increase of the shear rate but there is a shear thickening at a certain stage.

Moreover, we specifically explore the self-assembly of the block copolymer/homopolymer under a weak and strong shear in two different initial states, respectively, and obtain different results. The order parameter is in good agreement with this result. Interestingly, when the initial state is an ordered structure, our results are just in agreement with the conclusion under strong shear studied by L. L. He \cite{key47}. Under a weak shear, the domain structure merely tilts along with the shear flow. This is because the whole interfacial tension of the block copolymer/homopolymer is larger than the block copolymer/nanorod, and the order-order transition is more difficult under a weak shear. In other words, this is the effect of the homopolymer. Certainly, with regard to different results in different initial states, the thermal history also has a certain effect. Therefore, it is indispensable to apply the shear field at an appropriate time if we want to get what we want. Our result could provide a guideline for an experimentalist, and the model system can also give a simple way to realize an orientational order transition in soft materials through imposing a shear flow.

\newpage
\section*{Acknowledgements}
The project supported by the National Natural Science Foundation of China (Grant No. 21031003), the Specialized Research Fund for the Doctoral Program of Higher Education of China (Grant \linebreak No.~20121404110004 and No. 20091404120002), the Soft Science Program of Shanxi Province (Grant No. 2011041015-01), the Research Foundation for Excellent Talents of Shanxi Provincial Department of Human Resources and Social Security.

\newpage

\ukrainianpart

\title{Мікрофазні переходи блочний кополімер/гомополімер  за умов зсувової течії}

\author{Ю. Гуо, Дж. Шень, Б. Вень, Х. Ву, М. Сун, Дж. Пен}

\address{Школа хімії і матеріалознавства, Педагогічний університет Шаньсі, Ліньфень, 041004, Китай}

\makeukrtitle
\begin{abstract}

З метою дослідження  фазової поведінки суміші блочний кополімер/гомополімер під дією сталої зсувової течії використано моделювання динаміки комірок. Фазові переходи відбуваються від поперечної до паралельної, а потім до перпендикулярної ламеларної структури з ростом зсувової швидкості, що є результатом взаємодії між зсувовою течією і флуктуаціями концентрації. Усі реологічні властивості, такі як різниці нормального напруження і зсувова в'язкість, тісно пов'язані з напрямком ламели. Більше того, зокрема нами досліджено фазову поведінку і параметр порядку при слабкому і сильному зсувові двох різних початкових станів, з врахуванням при цьому важливості термічної історії. Для одержання бажаного результату зсувове поле слід прикласти у відповідний момент часу. Результати цієї роботи забезпечують  легкий спосіб створення впорядкованих бездефектних  матеріалів як в експерименті, так і в інженерній технології за рахунок введення  зсувової течії.

\keywords самоскупчення, блочний кополімер, гомополімер, зсувова течія
\end{abstract}

\end{document}